# Project Icarus: A Review of Local Interstellar Medium Properties of Relevance for Space Missions to the Nearest Stars

Ian A. Crawford*

*Department of Earth and Planetary Sciences, Birkbeck College, University of London, Malet Street, London, WC1E 7HX, UK*

*Tel: +44 207 679 3431; Email: i.crawford@ucl.ac.uk

**Abstract**

I review those properties of the interstellar medium within 15 light-years of the Sun which will be relevant for the planning of future rapid (v ≥ 0.1c) interstellar space missions to the nearest stars. As the detailed properties of the local interstellar medium (LISM) may only become apparent after interstellar probes have been able to make *in situ* measurements, the first such probes will have to be designed conservatively with respect to what can be learned about the LISM from the immediate environment of the Solar System. It follows that studies of interstellar vehicles should assume the lowest plausible density when considering braking devices which rely on transferring momentum from the vehicle to the surrounding medium, but the highest plausible densities when considering possible damage caused by impact of the vehicle with interstellar material. Some suggestions for working values of these parameters are provided. This paper is a submission of the Project *Icarus* Study Group.

**Keywords:** Interstellar medium; Interstellar travel

## 1 Introduction

The growing realisation that planets are common companions of stars [1,2] has reinvigorated astronautical studies of how they might be explored using interstellar space probes. The history of Solar System exploration to-date shows us that spacecraft are required for the detailed study of planets, and it seems clear that we will eventually require spacecraft to make *in situ* studies of other planetary systems as well. The desirability of such direct investigation will become even more apparent if future astronomical observations should reveal spectral evidence for life on an apparently Earth-like planet orbiting a nearby star (e.g., [3]). Definitive proof of the existence of such life, and studies of its underlying biochemistry, cellular structure, ecological diversity and evolutionary history will almost certainly require *in situ* measurements to be made [4]. This in turn will require the transport of sophisticated scientific instruments across interstellar space.

This is the context of the *Icarus* project [5] recently initiated by the Tau Zero Foundation [6] in collaboration with the British Interplanetary Society, and tasked with designing an interstellar space vehicle capable of making *in situ* scientific investigations of nearby stars. The specific target star has not yet been selected, but the project's Terms of Reference call for a nuclear fusion-based propulsion system, a vehicle able to decelerate at its target star, and a total mission duration "not exceeding a century and ideally much sooner" [7]. These constraints imply cruise velocities of the order of 0.1c or higher (where c is the speed of light), and a maximum realistic range of about 15 light-years from the Solar System. Among other considerations, it follows that an understanding of the properties of the Local Interstellar Medium (LISM) within this distance range will be important for the design of interstellar vehicles such as *Icarus*, and this is the topic of the present paper.

Properties of the LISM which may impact on the design of an interstellar space vehicle such as *Icarus* include:

- The density of the interstellar gas along the proposed trajectory
- The ionisation state of the gas
- The density and size distribution of solid interstellar dust grains
- The strength of the interstellar magnetic field

Even by interstellar standards the LISM has a very low density, and its properties would probably only be of scientific interest for a mission such as *Icarus* but for two caveats: (i) the density and ionisation state will be important if a magsail or similar device (e.g. [8]) is adopted as a deceleration mechanism [or, equivalently, if an interstellar ramjet [9] were to be considered as a means of fuel acquisition]; and (ii) the interstellar dust component will be a potential hazard for any space vehicle moving through the LISM with velocities of the order of 0.1c.

In the years since the pioneering *Daedalus* study [10] our knowledge of the LISM has improved significantly (e.g. [11-15]). It is generally accepted that the Sun is currently located close to the boundary of a small (spatial extent ≤ 10 light-years); low density ($n_H$ ~ 0.1-0.2 cm$^{-3}$, where $n_H$ is the density of hydrogen nuclei), warm (T ~ 7500 K), partially ionised interstellar cloud known as the Local Interstellar Cloud (LIC). Whether the Sun lies just within, or just outside, the LIC is currently a matter of debate (e.g. 13-15). The LIC is only one of a several broadly similar interstellar clouds within a few light-years of the Sun: Redfield & Linsky [13] identify six within 15 light-years (4.6 parsecs). These clouds are immersed in the empty ($n_H \approx n_e$ ~ 0.005 cm$^{-3}$), ionised, and probably hot (T ~ 10$^6$ K; but see [16] for an alternative view) Local Bubble (LB) in the interstellar medium [11,17]. The LB extends for about 60-100 parsecs from the Sun in the galactic plane before denser interstellar clouds are encountered, while at high galactic latitudes the LB appears to be open, forming a chimney-like structure in the interstellar medium which extends into the galactic halo (e.g. [18] and references cited therein). As discussed by Welsh and Shelton [16], there is currently some debate regarding the properties of the LB. In particular, although still supported by a number of observations and arguments [17], the presence of a 10$^6$ K plasma immediately surrounding the local clouds has recently been questioned. In response, Welsh and Shelton [16] have tentatively put forward an alternative model in which the LISM clouds are surrounded by a cooler (T ~ 20,000 K), denser ($n_e$ ~ 0.04 cm$^{-3}$) photo-ionised gas.

The properties of the LIC, and other nearby clouds, have been determined by spectroscopic studies of interstellar absorption lines towards nearby stars, augmented in the case of the LIC by observations of interstellar matter entering the Solar System and interacting with the heliosphere (e.g. Frisch et al. [15]). Information determined by these techniques which is of relevance for the design of interstellar space vehicles such as *Icarus* is reviewed in the following sections.

## 2. The distribution of nearby clouds

Redfield and Linsky [13] have conducted the most comprehensive survey to-date of interstellar absorption lines towards nearby stars, and have produced a catalogue of low-density interstellar clouds within 50 light-years (15 parsecs) of the Sun. The six clouds identified within 15 light-years, and therefore relevant to the *Icarus* study, are listed in Table 1. Within this distance range there are approximately 56 stars, in 38 separate stellar systems. Probably the most authoritative recent compilation is the RECONS (Research Consortium on Nearby Stars) list of the one hundred nearest star systems [19]. Using the maps of the extent of nearby clouds on the sky provided by Redfield and Linsky [13], it is possible to identify the interstellar cloud(s) through which an interstellar space vehicle would have to travel to get to any of these stars. This information is given in Table 2.

The individual clouds will be separated by much lower density Local Bubble material and, based on their thorough analysis of interstellar lines towards 157 nearby stars, Redfield and Linsky [13] estimated a volume filling factor of between 6 and 19% for LISM clouds within 15 parsecs (50 light-years). This implies that stars are more likely to lie in low density inter-cloud (Local Bubble) material than within the clouds themselves, although there appears to be a greater concentration of clouds closer to the Sun (half of those identified lie within 5 parsecs). The question of whether a given target star lies within a cloud, or within the low density Local Bubble material, becomes important for interstellar space vehicles if deceleration systems are considered which rely on an interaction between the space vehicle and the surrounding interstellar medium [8].

Several of the brightest stars within 15 light-years of the Sun have had their total neutral interstellar hydrogen column densities directly measured by absorption line spectroscopy [13,20]. These are listed in Table 3. In these seven cases, and after allowance for additional components of ionised hydrogen (probably of the order of 25%, see below) and ionised and neutral helium (about 10%), we already know the total quantity of interstellar gas which would have to be traversed by a space vehicle designed to travel to these particular stars. If we knew the actual spatial density (i.e. atoms or ions cm$^{-3}$) of the clouds we could estimate the relative fraction of each sightline occupied by clouds in these particular cases, and whether or not the target star lies within a cloud or in lower density material. Table 3 gives upper and lower estimates for the percentage of the path-length occupied by cloud material, based on assumed lower and upper limits for the density of the LIC and G clouds (discussed in Section 3).

Some indication of whether a particular star lies within a cloud is given by whether or not an astrosphere (the equivalent of the heliosphere surrounding the Sun) can be detected in the hydrogen Lyman-α absorption line [20]. The detection of an astrosphere in Ly-α requires neutral hydrogen to surround the star, so would not be expected if the star was immersed in the low density, ionised, Local Bubble material. As can be seen from Table 3, astrospheres have been detected around most of the nearest stars for which data are available [20], which implies that most of these stars are surrounded by moderately dense interstellar material. The main exception is τ Cet, where an astrosphere has been searched for but not detected.

The internal structure of these clouds is also of potential importance for interstellar mission design. Interstellar absorption lines provide information on the integrated column density of absorbing material, but not spatial inhomogeneities within it. Information on spatial structure can be obtained by comparing the interstellar spectra of stars that are close together on the sky, or by observing temporal changes in the interstellar spectrum of a single star as the relative motions of the Sun and star cause the line-of-sight to probe different regions of the cloud. For denser, more distant, interstellar clouds there is now considerable evidence for significant density and/or ionisation structure on scales of tens or hundreds of astronomical units [21], which could be a potential problem for interstellar space vehicles passing through them. However, to-date there is no evidence for significant spatial inhomogenieties within the nearby low density LISM clouds, based on essentially identical absorption line spectra of nearby double stars (e.g. α Cen A/B and α CMa A/B), and 18 closely spaced background stars in the Hyades cluster [13,22]. Thus, although it is important to keep an open mind, current indications are that the physical properties of the LISM clouds are relatively homogeneous. Of course, ultimately, obtaining detailed knowledge of the internal structure of the local interstellar medium is one of the *scientific* issues that we would expect an interstellar space probe to address [4,23].

Clearly, once an interstellar space mission is seriously contemplated for a nearby star there will have to be a concerted astronomical campaign to determine as much information as possible about the foreground interstellar medium, and the presence or absence of a stellar astrosphere. We note that the majority of potential targets within 15 light-years are M dwarfs (Table 2), for which obtaining this information will be difficult owing to the faintness of the stars and the presence of many stellar lines in the star's own spectrum. Furthermore, detection of astrospheres around M stars is further hampered by their generally weak stellar winds, despite their propensity for flare activity. For example, Wood et al. [24] searched for, but failed to find, an astrosphere surrounding Proxima Cen, even though it presumably resides in the same interstellar environment as α Cen A/B. If an interstellar mission were to be planned for a nearby M dwarf star, perhaps because of the detection of a planetary system, then a detailed astronomical study of its interstellar and circumstellar spectrum would necessitate the use of very large telescopes, owing to the intrinsic faintness of these stars, and a detailed understanding of M star atmospheres in order to separate interstellar and stellar absorption lines. Adding M stars to the list of nearby stars studied by absorption line spectroscopy would also greatly help in better defining the three-dimensional boundaries of the local clouds (i.e. "astronephography" [25]), which will assist in the future planning of interstellar missions.

## 3. Physical properties of LISM clouds

Our only direct knowledge of the physical properties of the LISM comes from measurements of interstellar gas and dust entering the Solar System, allowing for whatever disturbances may occur as the material enters the heliosphere and becomes influenced by the Sun's magnetic field and solar wind [e.g. 14,15,26,27]. Over recent years a consensus has emerged regarding the physical properties of the interstellar medium immediately adjacent to the heliosphere (sometimes referred to as the Circum-Heliospheric Interstellar Medium: CHISM), and these properties are listed in Table 4. Here, I have adopted the average of the hydrogen densities recommended by Bzowski et al. [28] ($n_{HI}$ = 0.18±0.03 $cm^{-3}$) and Frisch [14] ($n_{HI}$ = 0.195±0.005 $cm^{-3}$), but have retained the larger (~ 20%) errors recommended by Bzowski et al. [28]. The CHISM electron density ($n_e$ = 0.07±0.02 $cm^{-3}$) and temperature (T = 6300±400 K) are taken from Slavin and Frisch [29] and Frisch [14], and references cited by these authors. Within the quoted errors, these values are also consistent with early models which reproduce measurements of interstellar energetic neutral atoms in the Earth's vicinity by the *Interstellar Boundary Explorer* (*IBEX*) spacecraft (with the caveat that these models adopt a neutral hydrogen density at the lower end of the range: i.e. $n_{HI}$ = 0.15 cm$^{-3}$; c.f. Table 1 of Frisch et al. [30]).

For completeness, Table 4 also lists the CHISM magnetic field strength (B); there are a number of estimates for this value in the literature, and Table 4 gives two values. By assuming that magnetic and thermal pressures are balanced, Slavin and Frisch [29] obtained B~2.7μG, which is similar to previous estimates and comparable to astronomical measurements of typical magnetic field strengths (summarised by Frisch [14]). However, using Voyager 2 measurements of the deflection of the solar wind in the heliosheath, Opher et al. [31] have recently argued that the local interstellar magnetic field strength actually lies in the range 3.7 to 5.5 μG which, if confirmed, would imply that the magnetic energy density dominates the overall LIC energy budget.

Uncertainty remains as whether the CHISM properties represent the properties of the LIC or, as argued by Redfield and Linsky [13], the properties of a narrow transition zone between the LIC and G clouds and therefore an average of the two. Estimates of the properties of these two clouds, insofar as they have been determined by astronomical techniques independently of the CHISM, are also given in Table 4. It is possible to obtain a lower limit for the hydrogen density in a cloud towards a given star by dividing the observed HI column density by the stellar distance (i.e. by assuming that the cloud extends all the way from the Sun to the star). Based on the observed column density towards α Cen this yields a rigorous lower limit for the G cloud density of $n_{HI} \geq 0.1$ cm$^{-3}$, while the observed column densities towards ε Ind and τ Cet yield a similar limit for the LIC of $n_{HI} \geq 0.09$ cm$^{-3}$ (although the lack of an observed astrosphere for τ Cet may mean that the lower limit should be somewhat higher in this case). These lower limits are only half as large as the value inferred for the CHISM ($n_{HI} \sim 0.2$ cm$^{-3}$) which, in the absence of any other information, should probably be taken as the 'best guess' value for the LIC.

The electron density of the LIC has been determined by Redfield and Falcon [32] to be 0.12±0.04 cm$^{-3}$, based on absorption line measurements of excited singly ionized carbon (CII*) towards nearby stars. Redfield and Falcon do not give any corresponding value for the G cloud, but for the whole ensemble of LISM sightlines studied they find an electron density range of 0.06 to 0.21 cm$^{-3}$, and it seems likely that the electron density of the G cloud (and other discrete nearby clouds listed in Table 1) lie within this range. Note that the well-characterized LIC value is significantly higher than the CHISM value estimated independently, and for the same neutral hydrogen density would imply a somewhat higher ionisation fraction than estimated for the CHISM (i.e. $\chi_H \sim 0.39\pm0.13$ rather than $0.27\pm0.10$).

The last column of Table 4 is an attempt to give values for these parameters suitable for adoption by the *Icarus* design team. They are based on the preceding values in the table, but are intended to be conservative in the sense that, given current knowledge, the actual values almost certainly lie within the ranges given. Note that the values given in Table 4 relate to the properties of interstellar clouds through which an interstellar vehicle would have to pass. They do not necessarily represent the conditions at the target star itself; if the star does not lie within a cloud (which may be determinable from the presence or absence of an astrosphere; Table 3), the physical parameters would be those appropriate for the Local Bubble. As already noted, these are still uncertain: the long-standing model of a hot, empty LB ($n_{H\,tot} \approx n_e \approx 0.005$ cm$^{-3}$; $T \approx 10^6$ K [11]) while probably still viable (see Shelton [17]) has recently been questioned, and it is possible that the spaces between the LISM clouds may instead be filled with a cooler (T~20,000 K), higher density ($n_{H\,tot} \approx n_e \approx 0.04$ cm$^{-3}$) medium (Welsh and Sheldon [16]).

Ultimately, obtaining definitive values for many of these parameters, and their spatial variation, will be a scientific objective of the first interstellar space probes [4,23]. As far as *Icarus* is concerned, it would be wise to plan conservatively – that is, assume the lowest likely density if considering breaking devices which rely on transferring momentum from the vehicle to the surrounding medium, and the highest plausible densities when considering possible damage caused by impact of the vehicle with interstellar material.

**4. Interstellar dust**

In addition to the gaseous component of the interstellar medium, there is a solid component in the form of small (generally sub-micron-sized) solid particles, referred to as interstellar dust grains. As far as can be determined by astronomical observations, in most phases of the interstellar medium there is an approximately (but only approximately) constant ratio of ~100 between the mass of interstellar gas and the mass of dust in a given volume (e.g. Spitzer [33]).

Impacts with interstellar grains will be potentially damaging for space vehicles travelling at a significant fraction of the speed of light, and this will need to be factored into the vehicle design. The issue of shielding an interstellar space probe from interstellar dust grains was considered in detail in the context of the *Daedalus* study by Martin [34]. Martin adopted beryllium as a potential shielding material (owing to its low density and relatively high specific heat capacity), and found that several kg per square meter of this material would be eroded from exposed surfaces over the course of a six light-year flight at a speed of 0.1c (with the exact amount depending on the density of interstellar material; see his Table 6). Longer durations, and/or higher speeds, would result in greater ablation of shield material.

The total mass of interstellar grains which will be intercepted by a space vehicle during its flight, and the probability of encountering catastrophically large particles which the dust shield may not be able to protect against, depends on, among other variables, the size distribution of grains along the planned trajectory. In the warm, low density environment of the LISM, interstellar grains are expected mostly to be much smaller than a micron in size [35,36]. This expectation has been confirmed by the direct detection of interstellar dust grains entering the Solar System by dust detectors on the *Ulysses*, *Galileo* and *Cassini* spacecraft, for which the mean mass of the impacting particles has been found to be $3\times10^{-16}$ kg, corresponding to a grain radius of 0.3μm for a 'silicate' density of 2500 kg m$^{-3}$ [27,37]. Until recently, it would have been expected that 1 μm would be an absolute upper-limit for the radius of interstellar grains likely to be encountered in the LISM. For reference, the mass of a 1 μm-radius grain of silicate composition is $10^{-14}$ kg, and its kinetic energy at 0.1c would be 4.5 J, which ought to be easily accommodated by the vehicle's dust shield.

However, from the point of view of interstellar spacecraft design there is a potentially worrying twist: the same spacecraft instruments which have identified the mean radii of interstellar grains entering the Solar System to be ~0.3 μm [27,37] have also identified a high-mass tail to the grain population, extending to at least $10^{-13}$ kg (i.e. 2 µm radius) and perhaps as high as $10^{-12}$ kg (4.5 µm radius). Allowing for these larger grains, Landgraf et al. [37] inferred a total dust mass density in the LIC (strictly the CHISM) of $6.2\times10^{-24}$ kg m$^{-3}$ (which is actually a lower limit, as the smallest population of interstellar grains, those with radii <0.1 µm, are deflected at the heliopause and never enter the Solar System). This estimate is currently being reassessed on the basis of spacecraft data collected since the original measurements were made [27], but as of mid-2010 this value is still the best available for the local interstellar dust density (Dr H. Krüger, personal communication, 2010). It is a factor of about two higher than might be expected based on the astronomically determined grain size distribution of Mathis et al. [35] for an assumed local hydrogen density of $n_H = 0.2$ cm$^{-3}$. Following Martin's analysis [34] we find that this interstellar dust density would be expected to erode of the order of 5 kg m$^{-2}$ of shielding material over a six light-year flight at 0.1c (or 20 kg m$^{-2}$ at 0.2c; c.f. his Table 6). Clearly these shielding masses are starting to become significant (although they might be ameliorated by a different choice of shielding material, or a different strategy for dealing with dust impacts; see below).

Even more worrying, from the point of view of interstellar spacecraft design, is the fact that the upper-bound to the size distribution of interstellar dust particles in the solar neighbourhood is not currently well constrained. Radar observations of meteors entering the Earth's atmosphere [27,38,39] appear to have identified an incoming population of very large interstellar grains, having masses > $3\times10^{-10}$ kg (corresponding to radii >30µm for a silicate density). The possibility of encountering even larger particles over the course of a voyage of several light-years cannot be discounted. Converting the interstellar grain fluxes based on the spacecraft measurements (shown in Fig. 11 of Landgraf et al. [37]) to spatial densities (assuming that these grains enter the Solar System with a velocity of 26 km s$^{-1}$) and crudely extrapolating to higher masses, implies a spatial density of 100-µm grains of $4\times10^{-17}$ m$^{-3}$. If the gradient of this distribution is adjusted to fit the radar results, then the density of these very large grains would be about two orders of magnitude higher. Thus, over a six light-year ($5.7\times10^{16}$ m) flight we might expect between 2 and 200 impacts per square metre with such large particles. A 100 µm grain would have a mass of order $10^{-8}$ kg and a kinetic energy at 0.1c of $4.5\times10^6$ J. This is equivalent to a

1 kg mass impacting at a speed of 3 km s$^{-1}$, and it is not clear that a simple ablation-type shield could adequately protect the vehicle against impacts of this magnitude. On the other hand, it may not be valid to extrapolate the grain size distribution to such large masses, not least because of the difficulty of reconciling the presence of such large solid particles in the LISM with constraints imposed by both the well-characterised extinction of starlight and the cosmic abundances of the elements [36]. Clearly, much more work needs to be done to better determine the upper-limit to the size distribution of interstellar dust grains in the local interstellar medium before it will be possible to finalise a dust protection system for an *Icarus*-style vehicle.

In the meantime it would be wise to assume the worst and plan conservatively. One possibility might be to investigate the potential of double-walled 'Whipple Shields' [40] rather than the solid dust shield considered for *Daedalus* [34]. An intriguing version of the former would be to use a fluid-filled cavity as a shield, which may be more effective at absorbing and re-distributing the energy of hyper-velocity impacts [41]. However, the efficacy of such solutions at quasi-relativistic velocities would need to be carefully assessed. Alternatively, it may be necessary to consider sophisticated sensing techniques (perhaps based on radar or lidar) to detect large in-coming grains, and active means to destroy or deflect them. Perhaps the simplest dust protection system, first suggested by Bond [42] in the context of the *Daedalus* study, would be for the spacecraft to be preceded by a fine cloud of small dust particles (ejected from the vehicle, and thus travelling at the same velocity but a small distance ahead), such that any in-coming large grains will be destroyed by collisions within this artificial dust cloud before they have a chance to reach the main vehicle. Proposals for interstellar space vehicles such as *Icarus* will need carefully to consider all these options.

One final caveat with regard to the distribution of interstellar dust along the path-lengths to possible target stars: over the distance scales of interest the gas and dust may not be well-mixed, and a 'standard' gas/dust density ratio (~100 by mass) may not apply. Relatively small grains (≤ 0.3 μm) are expected to couple to the interstellar gas on scales of ≤ 1 parsec (~3 light-years [27]), but this is already comparable with the sizes of the LISM clouds. Thus, perfect mixing between gas and dust in these clouds may not be expected even for the small grains. With regard to the largest grains discussed above, the distance scale for coupling to the gas-phase of the interstellar medium would be hundreds or thousands of light-years [27] and no such mixing can be expected in the LISM. Rather, the fluxes of these large particles in the solar vicinity may stem directly from their source regions (whatever these may be), with negligible interaction with the low density clouds of the LISM. In the context of planning interstellar missions such as *Icarus*, it follows that one should not assume a significant decrease in the dust impact rate as the vehicle passes from a relatively high density medium (e.g. the LIC) into the surrounding low-density LB material.

**5. The LISM towards particular stars of interest**

In this Section, we briefly summarise what is known regarding the interstellar path-lengths to the five stars listed in Table 3 for which adequate information is available, and which would potentially by high-priority targets for an interstellar space mission such as *Icarus*.

5.1 α Cen

The nearest star system, α Cen A/B (plus Proxima Cen), lies in the direction of the G Cloud on the sky, and exhibits interstellar absorption lines due to the presence of this cloud along the line-of-sight. The Sun does not lie within the G Cloud (instead lying either in the outer region of the LIC or within a narrow transition zone between the two clouds), so the empirically determined properties of the CHISM (Table 4) cannot automatically be taken to be appropriate for the G Cloud. The detection of an astrosphere around α Cen A/B implies that these stars are embedded in the G Cloud ([20]; Dr S. Redfield, personal communication, 2010). The apparent lack of an astrosphere around Proxima Cen probably has more to do with the weakness of the M star's stellar wind than the lack of surrounding interstellar material (Wood et al. [24]). If the G cloud density were as low as $n_H = 0.1$ cm$^{-3}$ then it could extend almost all the way from the boundary of the LIC to the star. On the other hand, if the density of the G cloud is as high as 0.2 cm$^{-3}$, as found for the CHISM (and thus arguably for the LIC), the boundary of the G Cloud would not be encountered until about halfway to the star. If this picture is confirmed, there may be sufficient interstellar density in the vicinity of α Cen A/B to entertain the use of a magsail-like deceleration mechanism for *Icarus*, but one would want confirmation of this scenario before committing to such an implementation.

5.2 ε Eri, ε Ind, and τ Cet

These three stars all line in the direction of the LIC on the sky, and exhibit corresponding absorption lines. There is no doubt that a significant fraction of the path-length towards all three stars lies within the LIC. For a density as low as $n_H = 0.1$ cm$^{-3}$ the LIC could extend most of the way (given the uncertainties, perhaps all the way) to ε Ind and τ Cet, although the apparent lack of an astrosphere around the latter star [20] argues against this. The low interstellar column density towards ε Eri probably excludes the possibility of the LIC extending all the way to the star under any plausible assumptions. This begs the question as to what causes the astrosphere detected around ε Eri if the LIC does not extend all the way to the star, as there is no spectral evidence for other clouds in the line-of-sight [13]. More research is needed on this topic. For example, ε Eri is surrounded by a circumstellar dust disk [43], and it would be important to determine whether or not circumstellar absorption lines might not arise in the outskirts of the disk rather than in an astrosphere originating through interaction with a surrounding interstellar cloud. The implications for the choice of *Icarus* deceleration system for this high-priority star could be significant.

Finally we note that, if the average density of the LIC is as high as 0.2 cm$^{-3}$, as implied by the CHISM data (Table 4), then the LIC can extend less than half way towards all three stars, consistent with the lack of detected astrosphere towards τ Cet [20], but adding the mystery of the origin of the astrosphere of ε Ind to that already noted for ε Eri. These discrepancies underscore the need for further work in mapping the extent and densities of the nearest clouds, and the need to be conservative in the *Icarus* design if the option of targeting stars which may not be surrounded by relatively dense interstellar material is to be kept open.

5.3 61 Cyg

The double star 61 Cyg does not lie within the LIC as mapped on the sky by Redfield and Linsky [13], but it is close to the boundary. There is spectral evidence for two clouds towards 61 Cyg, which Redfield and Linsky identified as the 'Aql' and 'Eri' clouds, but noted that the former might in fact be confused with the LIC. For densities in the range 0.1 to 0.2 cm$^{-3}$, these clouds can between them only occupy a relatively small fraction (certainly <60%) of the total path-length. The presence of an astrosphere [20] implies that the star is embedded in one of these clouds. If the line-of-sight misses the LIC this could be either of the 'Aur' or 'Eri' clouds, the properties of which are largely unknown. If instead, the line-of-sight has clipped the LIC, the implication is that the first part of the path to 61 Cyg is through the LIC, followed by a substantial fraction of low-density LB material, before encountering the 'Eri' cloud within which the star itself is likely located.

## 6. Conclusions

This review of the properties of the local interstellar medium has revealed several conclusions relevant to the design of quasi-relativistic (v ≥ 0.1c) interstellar space vehicles, such as that considered in the *Icarus* study:

(1) Many important properties of the LISM are still poorly defined, including the distribution of nearby clouds, their physical properties (especially density and ionisation state), the size distribution of interstellar dust grains, and the nature of the 'inter-cloud' material which will occupy a significant fraction of the path-length towards many possible target stars. Obtaining more robust estimates for these properties will be important before finalising the design of an interstellar space vehicle such as *Icarus*.

(2) Given these uncertainties, it is important that the first generation of interstellar space vehicles be designed *conservatively* with respect to key interstellar medium properties. That is, such studies should assume the lowest likely density if considering braking devices which rely on transferring momentum from the vehicle to the surrounding medium (such as magsails [8]), but the highest plausible densities when considering possible damage caused by impact of the vehicle with interstellar material. Some suggestions for working values of these parameters are given in Table 4.

(3) In order to maintain flexibility with regard to the choice of target star, it would be a mistake to assume that the target will necessarily lie within one of the relatively high density LISM clouds whose properties are summarised in Table 4. This is probably a safe assumption for α Cen (which appears to lie well within the G Cloud), but the situation regarding other possible target stars is uncertain. Thus, unless an early decision is made to target the α Cen system specifically, the conservative planning assumption must be that the target star lies within low-density LB material. On the basis of current knowledge, the conservative value of the electron density in this case is 0.005 cm$^{-3}$, which would have to be taken into account when considering deceleration options. In the context of the *Icarus* study, maximum flexibility with regard to choice of target star would be retained if deceleration were achieved with the same fusion-based propulsion system used to accelerate the vehicle, because in that case the properties of the surrounding interstellar medium are essentially irrelevant.

(4) *In situ* spacecraft detections of interstellar dust entering the Solar System imply a local interstellar dust mass density of ~$6.2 \times 10^{-24}$ kg m$^{-3}$ [37]. This is somewhat higher than previous estimates, and will have to be taken into account by the dust protection system of any quasi-relativistic (v ≥ 0.1c) interstellar space vehicle. The requirements of the dust protection system become more demanding for longer journeys and, especially, for higher velocities.

(5) Observations over the last decade have revealed an unexpected high-mass tail to the local interstellar grain size distribution. Individual particles with masses as high as $10^{-12}$ kg (4.5 μm radius) are almost certainly present. Moreover, radar detections of interstellar dust particles entering the Earth's atmosphere (Krüger and Grün [27] and references therein) imply a population of interstellar grains with masses > $3 \times 10^{-10}$ kg (corresponding to radii >30μm). While there are difficulties reconciling the presence of such large grains with other astronomical observations [36], a conservative planning assumption would be that particles as large as 100 μm radius ($10^{-8}$ kg), and perhaps larger, might be encountered in the course of a several light-year journey through the

LISM. The kinetic energy of such large particles striking an interstellar space vehicle with a relative velocity of 0.1c are considerable ($4.5\times10^6$ J), and some kind of active dust detection and mitigation system may need to be considered.

**Acknowledgements**

I thank Drs H. Krüger, S. Redfield, and B.Y Welsh, and for helpful correspondence regarding the structure of the LISM, Adam Crowl for drawing my attention to the *IBEX* results, and Andy Presby for pointing me to recent papers on the local interstellar magnetic field. I additionally thank Seth Redfield, Kelvin Long, and Rob Swinney for careful readings of the manuscript and for helpful comments which have improved it.

**References**

[1] S. Udry, N.C. Santos, Statistical properties of exoplanets, Ann. Rev. Astron. Astrophys. **45** (2007) 397-439.

[2] J. Schneider, The Extrasolar Planets Encyclopaedia, http://exoplanet.eu/ (accessed August 2010).

[3] C.S. Cockell, et al., Darwin—an experimental astronomy mission to search for extrasolar planets, Experimental Astronomy **23** (2009) 435-461.

[4] I.A. Crawford, The astronomical, astrobiological and planetary science case for interstellar spaceflight, J. Brit. Interplanet. Soc. **62** (2009) 415-421.

[5] K. Long, M. Fogg, R. Obousy, A. Tziolas, A. Mann, R. Osborne, A. Presby, Project Icarus: son of Daedalus – flying closer to another star, J. Brit. Interplanet. Soc. **62** (2009) 403-414.

[6] The Tau Zero Foundation, http://www.tauzero.aero/ (accessed August 2010).

[7] K. Long, Project Icarus: terms of reference document, http://www.icarusinterstellar.org/TOR.pdf (accessed August 2010).

[8] D. Andrews, R. Zubrin, Magnetic sails and interstellar travel', J. Brit. Interplanet. Soc. **43** (1990) 265-272.

[9] R.W. Bussard, Galactic matter and interstellar flight, Astronautica Acta **6** (1960) 179-194.

[10] A. Bond, A.R. Martin, R.A. Buckland, T.J. Grant, A.T. Lawton, H.R. Mattison, J.A. Parfit, R.C. Parkinson, G.R. Richards, J.G. Strong, G.M. Webb, A.G.A. White, P.P. Wright, Project Daedalus: Final Report, J. Brit. Interplanet. Soc. Suppl. (1978) S1-S192.

[11] D.P. Cox, R.J. Reynolds, The local interstellar medium, Ann. Rev. Astr. Astrophys. **25** (1987) 303-344.

[12] R. Lallement, P. Bertin, R. Ferlet, A. Vidal-Madjar, J.L. Bertaux, GHRS observations of Sirius-A I: interstellar clouds toward Sirius and local cloud ionization, *Astron. Astrophys*. **286** (1994) 898-908.

[13] S. Redfield, J.L. Linsky, The structure of the local interstellar medium, *Astrophys. J*. **673** (2008) 283-314.

[14] P.C. Frisch, Is the sun embedded in a typical interstellar cloud?, Space Sci. Rev. **143** (2009) 191-204.

[15] P.C. Frisch, et al., The galactic environment of the sun: interstellar material inside and outside of the heliosphere, Space Sci. Rev. **146** (2009) 235-273.

[16] B.Y. Welsh, R.L. Shelton, The trouble with the Local Bubble', Astrophys. Space Sci. **323** (2009) 1-16.

[17] R.L. Shelton, Revising the Local Bubble model due to solar wind charge exchange X-ray emission, Space Sci. Rev. **143** (2009) 231-239.

[18] I.A. Crawford, R. Lallement, R.J. Price, D.M Sfeir, B.P. Wakker, B.Y. Welsh, High-resolution observations of interstellar NaI and CaII towards the southern opening of the Local Interstellar Chimney: probing the disc-halo connection, Mon. Not. R. Astron. Soc. **337** (2002) 720-730.

[19] RECONS Group, The one hundred nearest star systems, http://www.chara.gsu.edu/RECONS/TOP100.posted.htm (accessed August 2010).

[20] B.E. Wood, S. Redfield, J.L. Linsky, H.-R. Müller, G.P. Zank, Stellar Ly-alpha emission lines in the Hubble Space Telescope archive: intrinsic line fluxes and absorption from the heliosphere and astrospheres, Astrophys. J. Suppl. **159** (2005) 118-140.

[21] I.A. Crawford, Variable interstellar absorption lines: a brief review, Astrophys. Space Sci. **285** (2003) 661-675.

[22] S. Redfield, J.L. Linsky, Microstructure of the Local Interstellar Cloud and identification of the Hyades Cloud, Astrophys. J. **551** (2001) 413-428.

[23] G.M. Webb, Project Daedalus: some principles for the design of a payload for a stellar flyby mission, in: A. Bond et al., Project Daedalus: Final Report, J. Brit. Interplanet. Soc. Suppl. (1978) S149-S161.

[24] B.E. Wood, J.L. Linsky, H.-R. Müller, G.P. Zank, Observational estimates for the mass-loss rates of α Centauri and Proxima Centauri using Hubble Space Telescope Lyα spectra, Astrophys. J. (2001) **547**, L49-L52.

[25] J.L. Linsky, S. Redfield, B.E. Wood, N. Piskunov, The three-dimensional structure of the warm local interstellar medium I: methodology, Astrophys. J. **528**, (2000) 756-766.

[26] J.D. Slavin, The origins and physical properties of the complex of local interstellar clouds, Space Sci. Rev. **143** (2009) 311-322.

[27] H. Krüger, E. Grün, Interstellar dust inside and outside the heliosphere, Space Sci. Rev. **143** (2009) 347-356.

[28] M. Bzowski, E. Möbius, S. Tarnopolski, V. Izmodenov, G. Gloeckler, Density of neutral interstellar hydrogen at the termination shock from Ulysses pickup ion observations, Astron. Astrophys. **491** (2008) 7-19.

[29] J.D. Slavin, P.C. Frisch, The boundary conditions of the heliosphere: photoionization modelscConstrained by interstellar and in situ data, Astron. Astrophys. **491** (2008) 53-68.

[30] P.C. Frisch et al., Can IBEX identify variations in the galactic environment of the sun using energetic neutral atoms?, Astrophys. J. **719** (2010) 1984-1992.

[31] M. Opher, F. Alouani Bibi, G. Toth, J. D. Richardson, V.V. Izmodenov, T. I. Gombosi, A strong, highly-tilted interstellar magnetic field near the Solar System, Nature **462** (2009) 1036-1038.

[32] S. Redfield, R.E. Falcon, The structure of the local interstellar medium V: electron densities, Astrophys. J. **683** (2008) 207-225.

[33] L. Spitzer, Physical Processes in the Interstellar Medium, John Wiley, New York, 1978.

[34] A.R. Martin, Project Daedalus: bombardment by interstellar material and its effects on the vehicle, in: A. Bond, et al., Project Daedalus Final Report, JBIS Suppl. (1978) S116-S121.

[35] J.S. Mathis, W. Rumpl, K.H. Nordsieck, The size distribution of interstellar grains, Astrophys. J. **217** (1977) 425-433.

[36] B.T. Draine, Perspectives on interstellar dust inside and outside of the heliosphere, Space Sci. Rev. **143** (2009) 333-345.

[37] M. Landgraf, W.J. Baggaley, E. Grün, H. Kruger, G. Linkert, Aspects of the mass distribution of interstellar dust grains in the solar system from in situ measurements, J. Geophys. Res. **105** (2000) 10343-10352.

[38] A.D. Taylor, W.J. Baggaley, D.I. Steel, Discovery of interstellar dust entering the Earth's atmosphere, Nature **380** (1996) 323-325.

[39] W.J. Baggaley, Advanced meteor orbit radar observations of interstellar meteoroids, J. Geophys. Res. **105** (2000) 10353-10362.

[40] F.L. Whipple, Meteorites and space travel, Astron. J. **52** (1947) 131-131.

[41] N.N. Smirnov, A.B. Kiselev, K.A. Kondratyev, S.N. Zolkin, Impact of debris particles on space structures modelling, Acta Astronautica **67** (2010) 333-343.

[42] A. Bond, Project Daedalus: target system encounter protection, in: A. Bond et al., Project Daedalus: Final Report, JBIS Suppl. (1978) S123-S125.

[43] D. Backman, M. Marengo, K. Stapelfeldt, K.Su, D. Wilner, C.D. Dowell, D. Watson, J. Stansberry, G. Rieke, T. Megeath, G. Fazio, M. Werner, Epsilon Eridani's planetary debris disk: structure and dynamics based on Spitzer and Caltech Submillimeter Observatory observations, Astrophys. J. **690** (2009) 1522-1538.

**TABLES**

**Table 1.** The six closest interstellar clouds to the Sun as identified by Redfield and Linsky [13]. Other than the acronym LIC ('Local Interstellar Cloud'), the names are informal adoptions by various investigators; $l$ and $b$ are the galactic longitude and latitude, respectively, of the cloud centre (but note they all have highly irregular shapes); the distance to the closest star which exhibits absorption lines due to particular cloud may be taken as an upper-limit to the distance of the near-face of the cloud.

| Cloud name (informal) | Dist. to closest star with cloud absorption (lt-yrs) | $l$ (deg) | $b$ (deg) | Area on Sky (square deg) |
|---|---|---|---|---|
| LIC | 8.5 | 170 | -10 | 18270 |
| 'G' | 4.4 | 315 | +00 | 8230 |
| 'Blue' | 8.5 | 250 | -30 | 2310 |
| 'Aql' | 11.4 | 40 | -05 | 2960 |
| 'Eri' | 11.4 | 70 | -20 | 1970 |
| 'Aur' | 11.4 | 210 | +10 | 1640 |

**Table 2.** List of star systems within 15 light-years of the Sun (RECONS, [19]), identifying those interstellar clouds present along each sight-line as determined by Redfield and Linsky [13]. A question mark indicates that a star is close to the boundary of a particular cloud on the sky (or, in the case of Ross 154 and the 'Aql' cloud, that the star may lie in front of the cloud).

| Dist. Order | GJ | Popular Name | Spectral Type | Distance (lt-yrs) | $l$ (deg) | $b$ (deg) | Cloud(s) |
|---|---|---|---|---|---|---|---|
| 1 | 551 | Proxima Cen | M5.5V | 4.2 | 313.9 | -01.9 | G |
|  | 559 | α Cen A<br>α Cen B | G2V<br>K0V | 4.4 | 315.7 | -00.7 | G |
| 2 | 699 | Barnard's Star | M4V | 6.0 | 031.0 | +14.1 | LIC?, G? |
| 3 | 406 | Wolf 359 | M6V | 7.8 | 244.1 | +56.1 |  |
| 4 | 411 | Lalande 21185 | M2V | 8.3 | 185.1 | +65.4 | LIC |
| 5 | 244 | α CMa A (Sirius)<br>α CMa B | A1V<br>DA2 | 8.6 | 227.2 | -08.9 | LIC, Blue |
| 6 | 65 | Luyten 726-8 A<br>Luyten 726-8 B | M5.5V<br>M6V | 8.7 | 175.5 | -75.7 | LIC |
| 7 | 729 | Ross 154 | M3.5V | 9.7 | 011.3 | -10.3 | G?, Aql? |
| 8 | 905 | Ross 248 | M5.5V | 10.3 | 110.0 | -16.9 | LIC |
| 9 | 144 | ε Eri | K2V | 10.5 | 195.8 | -48.1 | LIC |
| 10 | 887 | Lacaille 9352 | M1.5V | 10.7 | 005.1 | -66.0 | LIC |
| 11 | 447 | Ross 128 | M4V | 10.9 | 270.1 | +59.6 |  |
| 12 | 866 | EZ Aqr A<br>EZ Aqr B<br>EZ Aqr C | M5V<br>M?<br>M? | 11.3 | 047.1 | -57.0 | LIC |
| 13 | 280 | α CMi A(Procyon)<br>α CMi B | F5IV-V<br>DA | 11.4 | 213.7 | +13.0 | LIC, Aur |
| 14 | 820 | 61 Cyg A<br>61 Cyg B | K5V<br>K7V | 11.4 | 082.3 | -05.8 | LIC?<br>Aql, Eri |
| 15 | 725 | Struve 2398 A<br>Struve 2398 B | M3V<br>M3.5V | 11.5 | 089.3 | +24.2 | LIC |
| 16 | 15 | (A) GX And<br>(B) GQ And | M1.5V<br>M3.5V | 11.6 | 116.7 | -18.4 | LIC |
| 17 | 845 | ε Ind A<br>ε Ind B<br>ε Ind C | K5Ve<br>T1<br>T6 | 11.8 | 336.2 | -48.0 | LIC |
| 18 | 1111 | DX Can | M6.5V | 11.8 | 197.0 | +32.4 | LIC |
| 19 | 71 | τ Cet | G8V | 11.9 | 173.1 | -73.4 | LIC |
| 20 | 1061 |  | M5.5V | 12.0 | 251.9 | -52.9 | G?, Blue? |
| 21 | 54.1 | YZ Cet | M4.5V | 12.1 | 149.7 | -78.8 | LIC |
| 22 | 273 | Luyten's Star | M3.5V | 12.4 | 212.3 | +10.4 | LIC, Aur? |
| 23 |  | Teegarden's Star | M7V | 12.5 | 160.3 | -37.0 | LIC |
| 24 |  | SCR1845-6357 A<br>SCR1845-6357 B | M8.5V<br>T | 12.6 | 331.5 | -23.5 | G |
| 25 | 191 | Kapteyn's Star | M1.5V | 12.8 | 250.5 | -36.0 | Blue |
| 26 | 825 | AX Mic | M0V | 12.9 | 003.9 | -44.3 | LIC?, Aql? |
| 27 | 860 | Kruger 60 A | M3V | 13.1 | 104.7 | +00.0 | LIC |

| | | Kruger 60 B | M4V | | | | |
|---|---|---|---|---|---|---|---|
| 28 | | DEN J1048-3956 | M8.5V | 13.2 | 278.7 | +17.1 | G |
| 29 | 234 | Ross 614 A<br>Ross 614 B | M4.5V<br>M8V | 13.3 | 212.9 | -06.2 | LIC? |
| 30 | 628 | Wolf 1061 | M3.0V | 13.8 | 003.4 | +23.7 | G |
| 31 | 35 | Van Maanen's Star | DZ7 | 14.1 | 121.9 | -57.5 | |
| 32 | 1 | | M3V | 14.2 | 343.6 | -75.9 | LIC |
| 33 | 473 | Wolf 424 A<br>Wolf 424 B | M5.5V<br>M7V | 14.3 | 288.8 | +71.4 | |
| 34 | 83.1 | TZ Ari | M4.5 | 14.5 | 147.7 | -46.5 | LIC |
| 35 | 687 | | M3V | 14.8 | 098.6 | +32.0 | LIC |
| 36 | 3622 | LHS 292 | M6.5V | 14.8 | 261.0 | +41.3 | Aur? |
| 37 | 674 | | M3V | 14.8 | 343.0 | -06.8 | G |
| 38 | 1245 | V1581 Cyg A<br>V1581 Cyg B<br>V1581 Cyg C | M5.5V<br>M6.0V<br>M? | 14.8 | 078.9 | +08.5 | LIC?, G?, Aql |

**Table 3**. The seven stars within 15 light-years of the Sun for which the interstellar neutral hydrogen column density, N(HI), has been measured [13]. Also given is the average spatial density, $<n_{HI}>$, obtained by dividing the column density by the stellar distance. This would only equal the actual density if the entire path-length is within the cloud, and therefore provides a rigorous lower-limit to this value. Columns six and seven give upper and lower limits, respectively, for the percentage of each sightline occupied by cloud material, based on lower and upper limits to the likely density of the nearby clouds (as discussed in Section 3). The last column indicates whether an astrosphere has been detected around the star [20]; the detection of an astrosphere implies the presence of neutral hydrogen around the star and thus that the star may lie within the more distant of the clouds present in its sightline (see text for discussion).

| Star | Dist (Lt-yr) | Log N(HI) (cm$^{-2}$) | $<n_{HI}>$ (cm$^{-3}$) | Cloud(s) | Cloud% (max) | Cloud% (min) | Astrosphere detected? |
|---|---|---|---|---|---|---|---|
| α Cen | 4.4 | 17.6 | 0.10 | G | 100 | 42 | Yes |
| α CMa | 8.6 | 17.4 | 0.03 | LIC, Blue | 30 | 14 | |
| ε Eri | 10.5 | 17.8 | 0.06 | LIC | 60 | 28 | Yes |
| α CMi | 11.4 | 17.9 | 0.07 | LIC, Aur | 70 | 32 | |
| 61 Cyg | 11.4 | 17.8 | 0.06 | Aql, Eri | 60 | 26 | Yes |
| ε Ind | 11.8 | 18.0 | 0.09 | LIC | 90 | 39 | Yes |
| τ Cet | 11.8 | 18.0 | 0.09 | LIC | 90 | 39 | No |

**Table 4.** Physical properties if the CHISM, LIC, and G Cloud, compiled from various sources (see text). Here, $n_{HI}$ is the spatial density of neutral atomic hydrogen; $n_e$ is the electron density; $n_{H\ tot}$ is the total hydrogen density (neutral plus ionised); $\chi_H$ is the hydrogen ionisation fraction; T is the kinetic temperature of the gas; and B is the local interstellar magnetic field strength (see text for explanation of the two values given).

| Parameter | CHISM | LIC | G Cloud | Suggested conservative range for *Icarus* design study* |
|---|---|---|---|---|
| $n_{HI}$ (cm$^{-3}$) | 0.19±0.04 | ≥ 0.09 | ≥ 0.10 | 0.10 – 0.23 |
| $n_e$ (cm$^{-3}$) | 0.07±0.02 | 0.12±0.04 | 0.06 – 0.21 ? | 0.05 – 0.21 |
| $n_{H\ tot}$ (cm$^{-3}$) | 0.26±0.05 | ≥ 0.17 | ≥ 0.16 | 0.15 – 0.43 |
| $\chi_H$ | 0.27±0.10 | ≤ 0.64 | ≤ 0.68 | 0.17 – 0.50 |
| T (K) | 6300±400 | 7500±1300 | 5500±400 | 5000 – 9000 |
| B (µG) | 2.7; 3.7–5.5 | N/A | N/A | 2.7 – 5.5 |

*Note that these values relate to the properties of interstellar clouds through which the *Icarus* vehicle will have to pass. They do not necessarily represent the conditions along the entire path to a given star, or those at the target star itself. If the star does not lie within a cloud the physical parameters would be those appropriate for the lower density Local Bubble (corresponding conservative ranges for which might be: $n_{H\ tot} \approx n_e$: 0.005 – 0.04 cm$^{-3}$; T: $2\times10^4 - 10^6$ K).